\documentclass[prb,twocolumn,floatfix,showpacs,amsmath,amssymb,superscriptaddress]{revtex4}

\usepackage{graphicx}
\usepackage{dcolumn}
\usepackage{bm}
\usepackage{hyperref}
\usepackage[latin1]{inputenc}

\begin{document}

\title{Generalized scattering-matrix approach for magneto-optics in periodically
patterned multilayer systems}

\author{B. Caballero}\affiliation{IMM-Instituto de
Microelectr\'onica de Madrid (CNM-CSIC), Isaac Newton 8, PTM, Tres
Cantos, E-28760 Madrid, Spain} \affiliation{Departamento de
F\'{\i}sica Te\'orica de la Materia Condensada, Universidad
Aut\'onoma de Madrid, 28049 Madrid, Spain.}

\author{A. Garc\'{\i}a-Mart\'{\i}n}
\affiliation{IMM-Instituto de Microelectr\'onica de Madrid
(CNM-CSIC), Isaac Newton 8, PTM, Tres Cantos, E-28760 Madrid,
Spain}

\author{J. C. Cuevas}
\affiliation{Departamento de F\'{\i}sica Te\'orica de la Materia Condensada,
Universidad Aut\'onoma de Madrid, 28049 Madrid, Spain.}
\affiliation{Department of Physics, University of Konstanz, 78457 Konstanz, Germany.}

\date{\today}

\begin{abstract}
We present here a generalization of the scattering-matrix approach
for the description of the propagation of electromagnetic waves in
nanostructured magneto-optical systems. Our formalism allows us to
describe all the key magneto-optical effects in any configuration
in periodically patterned multilayer structures. The method can
also be applied to describe periodic multilayer systems comprising
materials with any type of optical anisotropy. We illustrate the
method with the analysis of a recent experiment in which the
transverse magneto-optical Kerr effect was measured in a Fe film
with a periodic array of subwavelength circular holes. We show, in
agreement with the experiments, that the excitation of surface
plasmon polaritons in this system leads to a resonant enhancement
of the transverse magneto-optical Kerr effect.
\end{abstract}


\maketitle

\section{Introduction}

In recent years a lot of attention has been paid to the study of
the optical properties of nanostructured materials with both
plasmonic and magneto-optic activity.\cite{Armelles2009} The key
idea is to use hybrid nanostructures containing both metals, which
exhibit plasmon resonances, and ferromagnetic materials, which
provide high magneto-optical activity for reasonably low values of
the applied magnetic field, to profit from the best of the worlds
of plasmonics and magneto-optics. In the context of these hybrid 
structures there are two main topics of interest. The first one is
the use of the localization of the electromagnetic field due the
excitation of the plasmons supported by the free electrons in
metals to enhance the magneto-optical signals (Kerr effect,
Faraday effect, etc.).\cite{Hermann2001,Gonzalez2007} The
nanostructuring in these magneto-plasmonic structures plays a
fundamental role for several reasons. First of all, it provides a
convenient way to couple the light of an external source to the
plasmons supported by these hybrid systems, avoiding so the
typical wave vector mismatch in unstructured systems. On the
other hand, by nanostructuring these hybrid systems one can
manipulate light at the nanometer scale in several ways. In
particular, the enhancement of the electromagnetic field can be
largely increased since one can concentrate light in reduced
volumes. This can be done either by making use of localized plasmon 
resonances in isolated structures such as wires,\cite{Melle2003,
Gonzalez2007} disks\cite{Gonzalez2008,Gonzalez2010,Meneses2011,Chen2011},
or particles,\cite{Jain2009,Wang2011} or by a periodic perforation
in an otherwise continuous film.\cite{Diwekar2004,Papaioannou2010,
Ctistis2009,Torrado2010,Torrado2011,Papaioannou2011}

A second topic of interest is the use of magneto-optical effects to 
externally control either the properties of the transmission through perforated
membranes\cite{Battula2007,Strelniker2008,Belotelov2007,Wurtz2008}
or the very value of the surface plasmon wave vector.\cite{Gonzalez2007,
Temnov2010,Martin-Becerra2010,Torrado2010b,Belotelov2011} In this latter 
case, the relevant configuration is the transverse magneto-optical Kerr 
effect (see Sec.~\ref{sec-antidot} below) since the other configurations 
induce a polarization conversion that implies a decoupling of the plasmon.

In view of the relevance of these novel hybrid structures, and in
order to guide their design, it is crucial to have theoretical
methods that are able to describe the wave propagation in
nanostructured magneto-plasmonic systems. A powerful approach,
which is widely used to describe nanostructured systems without
magneto-optical activity, is the so-called scattering-matrix
formalism.\cite{Whittaker1999,Tikhodeev2002,Anttu2011} In recent
years, this approach has been extended to study different
magneto-optical effects in nanostructured multilayer
structures\cite{Garcia-Martin2005} and to describe the wave
propagation in periodic structures containing certain types of
anisotropic media.\cite{Liscidini2008} However, there are still
basic physical situations which lie out of the scope of the
existent implementations of the scattering formalism. Thus for
instance, the Kerr and the Faraday effects in the transverse
configuration, in which the magnetic field (or the magnetization
of the sample) is parallel to the sample plane but perpendicular
to the plane of incidence, cannot be addressed with the existent
scattering-matrix-based approaches. More generally, when the
optical anisotropy of the the materials involves off-diagonal
elements of the permittivity tensor along the growth direction of
the multilayer structure, none of the existing implementations of
the scattering-matrix approach can describe the wave propagation
in such structures. The technical problem lies in the fact that in
such situations the propagating eigenstates in the different
layers cannot be simply described with a standard eigenvalue
problem. In this work, we show how this problem can be solved and
present a generalization of the scattering-matrix approach to
describe the magneto-optics of hybrid nanostructured systems in
any configuration. Moreover, the method can be applied to
periodically patterned multilayer systems comprising any kind of
optically anisotropic materials. We illustrate the capabilities of
our formalism by addressing a recent experiment in which the
transverse magneto-optical Kerr effect (TMOKE) was measured in a
periodically perforated Fe film.\cite{Torrado2010} We show that,
in  agreement with the experimental results, the excitation of
surface plasmon polaritons in these structures leads to an
enhancement of the TMOKE signal. More importantly, our theoretical
method paves the way for studying the interplay between
plasmon-driven effects and magneto-optics in a wide variety of
hybrid nanostructures.

The rest of the paper is organized as follows. In section \ref{SMA} we 
explain in detail how the scattering-matrix approach can be generalized 
to describe the wave propagation in all kind of magneto-optical and 
optically anisotropic periodic multilayer systems. Section \ref{sec-antidot} 
is devoted to the analysis of the experiments of Ref.~\onlinecite{Torrado2010}, 
which allows us to illustrate the power of the method. Then, we shall 
summarize the main conclusions of our work in section \ref{sec-conclusions}.
Finally, several technical issues related to the formalism developed in 
section \ref{SMA} are addressed in detail in three appendixes.

\section{Generalized scattering-matrix approach} \label{SMA}

Our central goal is to solve the Maxwell's equations for a patterned multilayer
structure containing any combination of materials (isotropic and anisotropic). For
this purpose, we shall generalize the scattering-matrix approach developed by
Whittaker and Culshaw in Ref.~\onlinecite{Whittaker1999}. Following this work,
we shall first discuss the Maxwell's equations to be solved. Then, we shall address
the band structure of an unbounded layer to determine the propagating eigenstates
in the different layers. Then, we shall discuss how to construct the fields in a
multilayer structure using those eigenstates and, finally we shall describe how
the scattering matrix can be used to determine the field amplitudes in the whole
structure.

\subsection{Maxwell's equations}\label{subsec-maxwell}

Let us start by describing the Maxwell's equations to be solved. Assuming a harmonic
time dependence $\exp(-i\omega t)$, the Maxwell's equations adopt the following
form: ${\bf \nabla} \cdot \epsilon_0 \bar \epsilon {\bf E} = 0$, ${\bf \nabla} \cdot
{\bf H} = 0$, ${\bf \nabla} \times {\bf H} = -i \omega \epsilon_0 \bar \epsilon {\bf E}$,
and ${\bf \nabla} \times {\bf E} =  i \omega \mu_0 {\bf H}$, where the permittivity
is in general a tensor given by
\begin{equation}
\label{perm-tensor}
\bar \epsilon = \left( \begin{array}{ccc}
\epsilon_{xx} & \epsilon_{xy} & \epsilon_{xz} \\
\epsilon_{yx} & \epsilon_{yy} & \epsilon_{yz} \\
\epsilon_{zx} & \epsilon_{zy} & \epsilon_{zz} \end{array} \right) .
\end{equation}
Notice that we have assumed that $\bar \mu= \bar 1$ since we are interested in the
optical regime. Notice also that for harmonic fields, the first Maxwell's equation
is implied by the third one and the second can be satisfied by expanding the magnetic
field in basis states with zero divergence. We now introduce the following rescaling:
$\omega \epsilon_0 {\bf E} \to {\bf E}$ and $\sqrt{\mu_0 \epsilon_0} \omega = \omega/c
\to \omega$. Thus, the final two equations to be solved are
\begin{eqnarray}
\label{Meq1}
{\bf \nabla} \times {\bf H} & = & -i \bar \epsilon {\bf E} , \\
\label{Meq2}
{\bf \nabla} \times {\bf E} & = & i \omega^2 {\bf H} .
\end{eqnarray}

We consider here multilayer systems in which each layer can be, in principle,
periodically structured. Thus, the tensor $\bar \epsilon$ is
independent of $z$, where $z$ corresponds to the growth direction of the structure,
and it depends on the in-plane coordinates ${\bf r} \equiv (x,y)$ in a periodic
fashion. Due to this periodicity, it is convenient to work in a momentum representation
for the in-plane coordinates. Thus, for a given Bloch wave vector ${\bf k}$, we
can expand the fields as a sum over reciprocal lattice vectors ${\bf G}$
\begin{eqnarray}
{\bf H} ({\bf r}, z) & = & \sum_{\bf G} {\bf \tilde H}_{\bf k} ({\bf G}, z)
e^{i({\bf k} + {\bf G}) \cdot {\bf r}} , \\
{\bf E} ({\bf r}, z) & = & \sum_{\bf G} {\bf \tilde E}_{\bf k} ({\bf G}, z)
e^{i({\bf k} + {\bf G}) \cdot {\bf r}} ,
\end{eqnarray}
Following Ref.~\onlinecite{Whittaker1999}, we define the Fourier space vectors
\begin{eqnarray}
\boldsymbol{h}(z) & \equiv & \left[ {\bf \tilde H}_{\bf k} ({\bf G}_1, z) ,
{\bf \tilde H}_{\bf k} ({\bf G}_2, z), \dots \right]^{T} \\
\boldsymbol{e}(z) & \equiv & \left[ {\bf \tilde E}_{\bf k} ({\bf G}_1, z) ,
{\bf \tilde E}_{\bf k} ({\bf G}_2, z), \dots \right]^{T} .
\end{eqnarray}
Note that, although ${\bf \tilde H}_{\bf k}$ and ${\bf \tilde E}_{\bf k}$
depend on ${\bf k}$, the whole calculation is carried out for a fixed ${\bf k}$,
so such labels will be omitted in other symbols.

In what follows, we shall need the Fourier expansion of the components of the
permittivity tensor for the different layers
\begin{equation}
\label{epG-Fourier}
\tilde \epsilon_{ij} ({\bf G}) = \frac{1}{S} \int_{\rm unit \; cell} d{\bf r} \,
\epsilon_{ij}({\bf r}) e^{-i {\bf G} \cdot {\bf r}} ,
\end{equation}
where $i,j=x,y,z$, $S$ is the area of the in-plane unit cell, and the matrix $\hat
\epsilon_{ij}$ is such that $(\hat \epsilon_{ij})_{\bf G G^{\prime}} = \tilde
\epsilon_{ij}({\bf G - G^{\prime}})$. Analogously, the components of the index
tensor $\eta_{ij}({\bf r}) = [ \bar \epsilon^{-1}({\bf r})]_{ij}$ have Fourier
expansions $\tilde \eta_{ij}({\bf G})$ and matrix representations $\hat \eta_{ij}$.

With the notation just introduced, the momentum representation of a product such as
$\epsilon_{ij} {\bf E}$ becomes $\hat \epsilon_{ij} \boldsymbol{e}$. Thus, the
relevant Maxwell's equations, Eqs.~(\ref{Meq1}) and (\ref{Meq2}), can be written,
in component form, as
\begin{eqnarray}
\label{Amp1}
i\hat k_y h_z(z) - h^{\prime}_y(z) & = & -i \sum_j \hat \epsilon_{xj} e_j(z) \\
\label{Amp2}
h^{\prime}_x(z) - i\hat k_x h_z(z) & = & -i \sum_j \hat \epsilon_{yj} e_j(z) \\
\label{Amp3}
i\hat k_x h_y(z) - i \hat k_y h_x(z) & = & -i \sum_j \hat \epsilon_{zj} e_j(z) ,
\end{eqnarray}
and
\begin{eqnarray}
\label{Farad1}
i\hat k_y e_z(z) - e^{\prime}_y(z) & = & i \omega^2 h_x(z) \\
\label{Farad2}
e^{\prime}_x(z) - i\hat k_x e_z(z) & = & i \omega^2 h_y(z) \\
\label{Farad3}
i\hat k_x e_y(z) - i\hat k_y e_x(z) & = & i \omega^2 h_z(z) ,
\end{eqnarray}
where $\hat k_x$ and $\hat k_y$ are diagonal matrices with $(\hat k_x)_{\bf
G G} = (k_x + G_x)$ and $(\hat k_y)_{\bf G G} = (k_y + G_y)$, and the primes
denote differentiation with respect to $z$.

To conclude this subsection, let us say that matrices like $\hat \epsilon_{ij}$ or
$\hat \eta_{ij}$ have in practice a finite dimension equal to $N_G \times N_G$, where
$N_G$ is number of reciprocal lattice vectors considered in the numerical calculations.
It is also worth stressing that the simple Fourier factorization used above for the
products like $\epsilon_{ij} E_j$, which is exact when $N_G \to \infty$, may lead
in some cases to serious convergence problems when truncating the matrices
$\hat \epsilon_{ij}$. The reason is that both the permittivity tensor and the
electric field can exhibit discontinuities at the interfaces between different
materials. The correct Fourier factorization of this type of products when $N_G$
is finite is discussed in detail in Appendix A.

\subsection{Band structure of a single layer}

Now our task is to solve the Maxwell's equations in momentum space derived in the
previous subsection for the case of an unbounded layer. In this case, the fields
have a $z$ dependence typical from plane waves, {\it i.e.}\ $\exp(iqz)$. Moreover,
the ${\bf H}$ field will be expanded in basis set of zero divergence, to satisfy
${\bf \nabla} \cdot {\bf H} = 0$, and the coefficients in this expansion will
be determined by substituting into Maxwell's equations.

Following Ref.~\onlinecite{Whittaker1999}, the magnetic field is expanded in
terms of $z$ propagating plane waves as follows
\begin{eqnarray}
{\bf H}({\bf r},z) & = & \sum_{\bf G} \left( \tilde \phi_x({\bf G}) \left[
{\bf \hat x} - \frac{1}{q} (k_x + G_x) {\bf \hat z} \right] \right. \\
& & \left. + \tilde \phi_y({\bf G}) \left[ {\bf \hat y} - \frac{1}{q} (k_y + G_y)
{\bf \hat z} \right] \right) e^{i({\bf k} + {\bf G}) \cdot {\bf r} + iqz} , \nonumber
\end{eqnarray}
where ${\bf \hat x}$, ${\bf \hat y}$, and ${\bf \hat z}$ are the Cartesian unit
vectors and $\tilde \phi_x({\bf G})$ and $\tilde \phi_y({\bf G})$ are the expansion coefficients
to be determined. Notice that this expression satisfies ${\bf \nabla} \cdot {\bf H}
= 0$. Now, it is convenient to rewrite the previous expression in momentum
representation. By defining the vectors $\phi_x = [ \tilde \phi_x({\bf G_1}), \tilde
\phi_x({\bf G_2}), \dots ]^{T}$ and $\phi_y = [ \tilde \phi_y({\bf G_1}), \tilde
\phi_y({\bf G_2}), \dots ]^{T}$, we can write
\begin{equation}
\boldsymbol{h}(z) = e^{iqz} \left\{ \phi_x {\bf \hat x} + \phi_y {\bf \hat y} -
\frac{1}{q} (\hat k_x \phi_x + \hat k_y \phi_y ) {\bf \hat z} \right\} ,
\end{equation}
where $\hat k_x$ and $\hat k_y$ are the diagonal matrices defined in Sec.~II.
For what follows, it is convenient to rewrite this last equation in the following
vector notation
\begin{equation}
\boldsymbol{h}(z) = e^{iqz} \left( \phi_x, \phi_y , - \frac{1}{q}
(\hat k_x \phi_x + \hat k_y \phi_y ) \right)^T , \label{h-kspace}
\end{equation}
where let us recall that every entry in this column vector is a vector of dimension
$N_G$. With this vector notation, Eqs.~(\ref{Amp1}-\ref{Amp3}) can now be written as
\begin{equation}
{\cal C} \boldsymbol{h}(z) = \hat{\hat \epsilon} \boldsymbol{e}(z) ,
\label{Amp-matrix}
\end{equation}
where the block matrices ${\cal C}$ and $\hat{\hat \epsilon}$
\begin{equation}
{\cal C} = \left( \begin{array}{ccc}
\hat 0 & q \hat 1 & -\hat k_y \\
-q \hat 1 & \hat 0 & \hat k_x \\
\hat k_y & - \hat k_x & \hat 0 \end{array} \right), \;\;
\hat{\hat \epsilon} = \left( \begin{array}{ccc}
\hat \epsilon_{xx} & \hat \epsilon_{xy} & \hat \epsilon_{xz} \\
\hat \epsilon_{yx} & \hat \epsilon_{yy} & \hat \epsilon_{yz} \\
\hat \epsilon_{zx} & \hat \epsilon_{zy} & \hat \epsilon_{zz}  \end{array} \right) .
\end{equation}
On the other hand, Eqs.~(\ref{Farad1}-\ref{Farad3}) adopt now the form
\begin{equation}
{\cal C}^{T} \boldsymbol{e}(z) = \omega^2 \boldsymbol{h}(z) .
\label{Farad-matrix}
\end{equation}
From Eq.~(\ref{Amp-matrix}) we obtain the following expression for the electric
field in momentum representation
\begin{equation}
\label{e-kspace}
\boldsymbol{e}(z) = \hat{\hat \eta} {\cal C} \boldsymbol{h}(z) ,
\end{equation}
where $\hat{\hat \eta} = \hat{\hat \epsilon}^{-1}$. Substituting this expression in
Eq.~(\ref{Farad-matrix}) we obtain the following closed equation for the magnetic field
in momentum representation
\begin{equation}
{\cal C}^{T} \hat{\hat \eta} {\cal C} \boldsymbol{h}(z) =
\omega^2 \boldsymbol{h}(z) ,
\end{equation}
which defines an eigenvalue problem for $\omega^2$. Indeed, only two of the three
identities obtained from this equation, one for each ${\bf \hat x}$, ${\bf \hat y}$,
and ${\bf \hat z}$, are independent. From the first two identities, and using
Eq.~(\ref{h-kspace}), we obtain the following equations determining the allowed
values for $q$
\begin{equation}
\left( {\cal A}_2 q^2 + {\cal A}_1 q + {\cal A}_0 + {\cal A}_{-1} \frac{1}q
\right) \phi = 0 ,
\label{rational-eigen}
\end{equation}
where $\phi = (\phi_x, \phi_y)^{T}$ and the $2\times 2$ block matrices ${\cal A}_n$
are defined by
\begin{widetext}
\begin{eqnarray}
& & {\cal A}_2 = \left( \begin{array}{cc} \hat \eta_{yy} & -\hat \eta_{yx} \\
-\hat \eta_{xy} & \hat \eta_{xx} \end{array} \right) , \;\;
{\cal A}_1 = {\cal A}^{(a)}_1 + {\cal A}^{(b)}_1 =
\left( \begin{array}{cc} -\hat k_y \hat \eta_{zy} & \hat k_y \hat \eta_{zx} \\
\hat k_x \hat \eta_{zy} & -\hat k_x \hat \eta_{zx} \end{array} \right) +
\left( \begin{array}{cc} -\hat \eta_{yz} \hat k_y & \hat \eta_{yz} \hat k_x \\
\hat \eta_{xz} \hat k_y & -\hat \eta_{xz} \hat k_x \end{array} \right) , \nonumber \\
& & {\cal A}_0 = {\cal A}^{(a)}_0 + {\cal A}^{(b)}_0 -\omega^2 {\bf \hat 1} =
\left( \begin{array}{cc} \hat k_y \hat \eta_{zz} \hat k_y & -\hat k_y \eta_{zz} \hat k_x  \\
-\hat k_x \hat \eta_{zz} \hat k_y & \hat k_x \eta_{zz} \hat k_x \end{array} \right) +
\left( \begin{array}{cc} \eta_{yy} \hat k_x \hat k_x - \eta_{yx} \hat k_y \hat k_x &
\hat \eta_{yy} \hat k_x \hat k_y - \hat \eta_{yx} \hat k_y \hat k_y \\
\hat \eta_{xx} \hat k_y \hat k_x -\hat \eta_{xy} \hat k_x \hat k_x &
\hat \eta_{xx} \hat k_y \hat k_y -\hat \eta_{xy} \hat k_x \hat k_y \end{array}
\right) - \omega^2 \left( \begin{array}{cc} 1 & 0 \\ 0 & 1  \end{array} \right) ,
\nonumber \\
& & {\cal A}_{-1} = \left( \begin{array}{cc} \hat k_y \hat \eta_{zx} \hat k_y \hat k_x
-\hat k_y \hat \eta_{zy} \hat k_x \hat k_x & \hat k_y \hat \eta_{zx} \hat k_y \hat k_y
-\hat k_y \hat \eta_{zy} \hat k_x \hat k_y \\ \hat k_x \hat \eta_{zy} \hat k_x \hat k_x
- \hat k_x \hat \eta_{zx} \hat k_y \hat k_x & \hat k_x \hat \eta_{zy} \hat k_x \hat k_y
- \hat k_x \hat \eta_{zx} \hat k_y \hat k_y \end{array} \right) . \label{A-def}
\end{eqnarray}
\end{widetext}
In general, Eq.~(\ref{rational-eigen}) is a so-called rational eigenvalue problem. This
problem belongs to the category of nonlinear eigenvalue problems, which continues to
be a challenge in the field of numerical analysis. However, we have found that a simple
linearization strategy allows us to solve such an eigenvalue problem in all the examples
that we have studied. The details of this method are explained in Appendix B. It is worth
stressing that so far the scattering approach has only been applied to situations where
the materials are isotropic\cite{Whittaker1999} or in cases in which the magneto-optical
activity is such that the off-diagonal components of the permittivity tensor involving the
$z$ component are zero\cite{Garcia-Martin2005} ($\epsilon_{xz} = \epsilon_{yz} = 0$). In
those cases, Eq.~(\ref{rational-eigen}) reduces to
\begin{equation}
{\cal A}_0 \phi = - {\cal A}_2 q^2 \phi ,
\end{equation}
which is a generalized eigenvalue problem for $q^2$, which can be solved with standard
techniques of linear algebra. Notice that, as explained in the introduction, those cases
exclude, for instance, the analysis of the Kerr effect in the transversal configuration.

The solution of Eq.~(\ref{rational-eigen}) provides $4N_G$ non-vanishing complex
eigenvalues for $q$. Half of these eigenvalues lie in the upper half of the complex
plane and half of them in the lower half. Finally, let us say that in the case of
spatially uniform slabs, Eq.~(\ref{rational-eigen}) reduces to a quartic equation for
$q$, which is well-known in the context of wave propagation in anisotropic media. This
is shown in Appendix C.

\subsection{Electric and magnetic field}

The next step toward the complete solution of the Maxwell's equations in a multilayer
structure is the determination of the fields in the different layers, which can be
done by expressing them in terms of the propagating wave eigenstates defined in the
previous subsection. To be precise, the fields can be expressed as a combination of
forward and backward propagating waves with wave numbers $q_n$, and complex amplitudes
$a_n$ and $b_n$, respectively. These amplitudes will be later on determined by
using the boundary conditions at the interfaces and surfaces of the multilayer
structure. Since the boundary conditions are simply the continuity of the in-plane
field components, we shall focus here on the analysis of the field components $e_x$,
$e_y$, $h_x$, and $h_y$.

From the momentum representation of ${\bf H}$ in Eq.~(\ref{h-kspace}), the in-plane
components of $\boldsymbol{h}$ can be expanded in terms of propagating waves as follows
\begin{eqnarray}
\left( \begin{array}{c} h_x(z) \\ h_y(z) \end{array} \right) & = & \sum_n \left\{
\left( \begin{array}{c} \phi_{x_n} \\ \phi_{y_n} \end{array} \right) e^{iq_nz} a_n
\right. \nonumber \\ & & \hspace*{5mm} \left. + \left( \begin{array}{c} \varphi_{x_n} \\
\varphi_{y_n} \end{array} \right) e^{-ip_n(d-z)} b_n \right\} ,
\end{eqnarray}
where $d$ is the thickness of the layer. Here, $a_n$ is the coefficient of the forward
going wave at the $z=0$ interface, and $b_n$ is the backward going wave at $z=d$.
On the other hand, $q_n$ correspond to the eigenvalues of Eq.~(\ref{rational-eigen})
with $\mbox{Im} \{q_n\} > 0$ and $p_n$ are the eigenvalues with $\mbox{Im} \{p_n\}
< 0$. Notice also that, contrary to the case of isotropic materials, here the
eigenfunctions of the forward and backward propagating waves are different in general.

To make the notation more compact, we now define two $2N_G \times 2N_G$ matrices
$\Phi_+$ and $\Phi_-$ whose columns are the vectors $\phi_n$ and $\varphi_n$,
respectively. Moreover, we define the diagonal $2N_G \times 2N_G$ matrices
$\hat {\rm f}_+(z)$ and $\hat {\rm f}_-(d-z)$, such that $[\hat {\rm f}_+(z)]_{nn}
= e^{iq_nz}$ and $[\hat {\rm f}_-(d-z)]_{nn} = e^{-ip_n(d-z)}$, and the $2N_G$-dimensional
vectors $h_{||}(z) = [h_x(z), h_y(z)]^{T}$, $a = (a_1,a_2,\dots)^{T}$, and
$b = (b_1,b_2,\dots)^{T}$. In terms of these quantities, the in-plane
magnetic-field components become
\begin{equation}
h_{||}(z) = \Phi_+ \hat {\rm f}_+(z) a + \Phi_- \hat {\rm f}_-(d-z) b .
\label{hp-exp}
\end{equation}

Similarly, using the momentum representation of ${\bf E}$ from Eq.~(\ref{e-kspace})
it is straightforward to show that the in-plane components of the electric field,
$e_{||}(z) = [-e_y(z), e_x(z)]^{T}$ (note the skew), are given by
\begin{eqnarray}
e_{||}(z) & = & \left( {\cal A}^{(b)}_0 \Phi_+ \hat q^{-1} + {\cal A}^{(b)}_1 \Phi_+
+ {\cal A}_2 \Phi_+ \hat q \right) \hat {\rm f}_+(z) a \nonumber \\ & & \hspace*{-1.5cm}
+ \left( {\cal A}^{(b)}_0 \Phi_- \hat p^{-1} + {\cal A}^{(b)}_1 \Phi_-
+ {\cal A}_2 \Phi_- \hat p \right) \hat {\rm f}_-(d-z) b ,
\label{ep-exp}
\end{eqnarray}
where the ${\cal A}$'s are defined in Eq.~(\ref{A-def}) and we have defined the
$2N_G \times 2N_G$ diagonal matrices $\hat q$ and $\hat p$ such that $\hat q_{nn} =
q_n$ and $\hat p_{nn} = p_n$.

We can now combine Eq.~(\ref{hp-exp}) and (\ref{ep-exp}) into a single expression
as follows
\begin{eqnarray}
\label{M-def}
\left( \begin{array}{c} e_{||}(z) \\ h_{||}(z) \end{array} \right) & = & M \left(
\begin{array}{c} \hat {\rm f}_+(z) a \\ \hat {\rm f}_-(d-z) b \end{array} \right) \\
& = & \left( \begin{array}{cc} M_{11} & M_{12} \\ M_{21} & M_{22} \end{array} \right)
\left( \begin{array}{c} \hat {\rm f}_+(z) a \\ \hat {\rm f}_-(d-z) b \end{array} \right)
, \nonumber
\end{eqnarray}
where the $2N_G \times 2N_G$ matrices $M_{ij}$ are defined as
\begin{eqnarray}
M_{11} & = & {\cal A}^{(b)}_0 \Phi_+ \hat q^{-1} + {\cal A}^{(b)}_1 \Phi_+
+ {\cal A}_2 \Phi_+ \hat q , \nonumber \\
M_{12} & = & {\cal A}^{(b)}_0 \Phi_- \hat p^{-1} + {\cal A}^{(b)}_1 \Phi_-
+ {\cal A}_2 \Phi_- \hat p , \nonumber \\
M_{21} & = & \Phi_+ , \;\;\; M_{22} =  \Phi_- .
\end{eqnarray}

\subsection{The scattering matrix}

The final step in our calculation is to use the scattering matrix ($S$-matrix) to
compute the field amplitudes needed to describe the different relevant physical
quantities. This part of the calculation is practically independent of the type of
materials present in the structure (isotropic or anisotropic) and it is nicely explained
in section V of Ref.~\onlinecite{Whittaker1999}. We just include a brief discussion
here to make this work more self-contained.

By definition, the $S$-matrix relates the vectors of the amplitudes of forward and
backward going waves, $a_l$ and $b_l$, where $l$ now denotes the layer, in the
different layers of the structure as follows
\begin{equation}
\left( \begin{array}{c} a_l \\ b_{l^{\prime}} \end{array} \right) = S(l^{\prime},l)
\left( \begin{array}{c} a_{l^{\prime}} \\ b_l \end{array} \right) =
\left( \begin{array}{cc} S_{11} & S_{12} \\ S_{21} & S_{22} \end{array} \right)
\left( \begin{array}{c} a_{l^{\prime}} \\ b_l \end{array} \right) .
\label{S-def}
\end{equation}

The field amplitudes in two consecutive layers are related via the boundary
conditions for the fields, namely the continuity of the in-plane components
of the fields in every interface and surface. If we consider the interface
between the layer $l$ and the layer $l+1$, the corresponding boundary conditions
read
\begin{equation}
\left( \begin{array}{c} e_{||}(d_l) \\ h_{||}(d_l) \end{array} \right)_l =
\left( \begin{array}{c} e_{||}(0) \\ h_{||}(0) \end{array} \right)_{l+1} ,
\end{equation}
where $d_l$ is the thickness of layer $l$. From this condition, together with
Eq.~(\ref{M-def}), it is easy to show that the amplitudes in layers $l$ and
$l+1$ are related by the interface matrix $I(l,l+1) = M^{-1}_l M_{l+1}$ in
the following way
\begin{eqnarray}
\left( \begin{array}{c} \hat f_{l,+} a_l \\ b_l \end{array} \right) & = & I(l,l+1)
\left( \begin{array}{c} a_{l+1} \\ \hat f_{l+1,-} b_{l+1} \end{array} \right)
\nonumber \\ & = &
\left( \begin{array}{cc} I_{11} & I_{12} \\ I_{21} & I_{22} \end{array} \right)
\left( \begin{array}{c} a_{l+1} \\ \hat f_{l+1,-} b_{l+1} \end{array} \right) ,
\end{eqnarray}
where $\hat f_{l,+} = \hat {\rm f}_{l,+}(d_l)$ and $\hat f_{l+1,-} = \hat
{\rm f}_{l+1,-}(d_{l+1})$.

Now, with the help of the interface matrices, the $S$-matrix can be calculated
in an iterative way as follows. The matrix $S(l^{\prime},l+1)$ can be calculated
from $S(l^{\prime},l)$ using the definition of $S(l^{\prime},l)$ in Eq.~(\ref{S-def})
and the interface matrix $I(l,l+1)$. Eliminating $a_l$ and $b_l$ we obtain the relation
between $a_{l^{\prime}}$, $b_{l^{\prime}}$ and $a_{l+1}$, $b_{l+1}$, from which
$S(l^{\prime},l+1)$ can be constructed. This reasoning leads to the following
iterative relations
\begin{eqnarray}
S_{11}(l^{\prime},l+1) & = & \left[I_{11} - \hat f_{l,+} S_{12}(l^{\prime},l) I_{21}
\right]^{-1} \hat f_{l,+} S_{11}(l^{\prime},l) \nonumber \\
S_{12}(l^{\prime},l+1) & = & \left[I_{11} - \hat f_{l,+} S_{12}(l^{\prime},l) I_{21}
\right]^{-1} \nonumber \\ & & \times \left( \hat f_{l,+} S_{12}(l^{\prime},l) I_{22}
- I_{12} \right) \hat f_{l+1,-} \nonumber \\
S_{21}(l^{\prime},l+1) & = & S_{22}(l^{\prime},l) I_{21} S_{11}(l^{\prime},l+1) +
S_{21}(l^{\prime},l) \nonumber \\
S_{22}(l^{\prime},l+1) & = & S_{22}(l^{\prime},l) I_{21} S_{12}(l^{\prime},l+1) +
\nonumber \\ & & S_{22}(l^{\prime},l) I_{22} \hat f_{l+1,-} .
\end{eqnarray}
Starting from $S(l^{\prime},l^{\prime}) = 1$, one can apply the previous recursive
relations to a layer at a time to build up $S(l^{\prime},l)$.

From the knowledge of the $S$-matrix one can compute all the field amplitudes needed
to describe a physical situation. Thus for instance, labeling the surface $l=0$ and
the substrate $l=N$, the calculation of the reflectivity and the transmission
coefficients requires the knowledge of the amplitudes $b_0$ and $a_N$, which can
be calculated from $S(0,N)$. On the other hand, it may be interesting to calculate
the fields inside the structure, for which we need the amplitudes $a_l$ and $b_l$.
These can be obtained by calculating $S(0,l)$ and $S(l,N)$, and using Eq.~(\ref{S-def})
to get
\begin{eqnarray}
a_l & = & \left[1 - S_{12}(0,l) S_{21}(l,N) \right]^{-1} \nonumber \\
& & \times \left[ S_{11}(0,l) a_0 + S_{12}(0,l) S_{22}(l,N) b_N \right] \nonumber \\
b_l & = & \left[1 - S_{21}(l,N) S_{12}(0,l) \right]^{-1} \nonumber \\
& & \times \left[ S_{21}(l,N) S_{11}(0,l) a_0 + S_{22}(l,N) b_N \right] .
\end{eqnarray}

\section{TMOKE in perforated iron films} \label{sec-antidot}

In this section we shall illustrate the method just described by analyzing the
experiment reported in Ref.~\onlinecite{Torrado2010} in which the transverse
magneto-optical Kerr effect (TMOKE) was studied in a periodically perforated Fe 
film.  The TMOKE consists in an intensity change of the $p$-component of
the reflected light upon application of a magnetic field perpendicular to the 
plane of incidence of the light.\cite{Zvezdin1997} In the case of ferromagnetic 
materials, the magnetic field is used to reverse the magnetization ${\bf M}$
of the medium and the TMOKE is characterized by the following quantity
\begin{equation}
\label{def-Tmoke}
\mbox{TMOKE} = \frac{R_{pp}(+{\bf M}) - R_{pp}(-{\bf M})}{R_{pp}(+{\bf M}) +
R_{pp}(-{\bf M})} ,
\end{equation}
where $R_{pp}(\pm {\bf M})$ are the reflectivity along the $p$-channel for the two
opposite magnetizations, which in this configuration are perpendicular to the incidence
plane and parallel to the layers of the structure. As explained in the introduction,
this arrangement induces off-diagonal components of the permittivity tensor of the
ferromagnetic material in the $z$-direction (direction of the growth of the multilayer)
and therefore, it requires the use of the generalized scattering approach described
in the previous section.

\begin{figure}[t]
\begin{center}
\includegraphics[width=\columnwidth,clip]{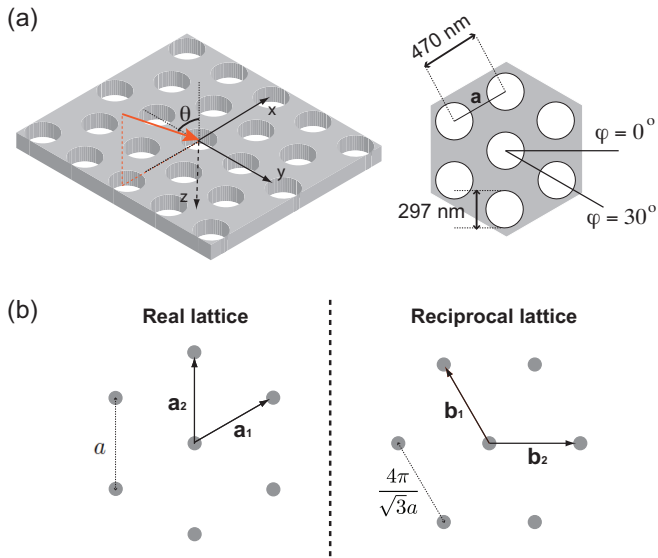}
\caption{(Color online) (a) Schematic representation of the system under study, which
consists of perforated Fe films with a periodic array of circular holes forming a
triangular lattice. Here, one can see the angle definitions and the geometrical
parameters of the hole array. (b) The triangular lattice both in real and in
reciprocal space, and definition of the basis vectors.}
\label{scheme}
\end{center}
\end{figure}

The structure studied in Ref.~\onlinecite{Torrado2010} is described 
schematically in Fig.~\ref{scheme}. It consists of a Fe film (100 nm 
thick), which is perforated with a periodic array of subwavelength circular 
holes (diameter of 297 nm) forming a triangular lattice with a lattice 
parameter of 470 nm. The Fe film was prepared on a Si(111) substrate and 
the structure contains additionally a seed layer of Ti (2 nm thick) and a 
capping layer of Au (2 nm thick), which were included to form a smooth Fe 
film and to prevent a subsequent oxidation of the surface, respectively. 
In our calculations we used the energy dependent permittivities taken 
from ellipsometric measurements of 20 nm-thick continuous films, 
and the off-diagonal elements of the ferromagnetic material have been 
extracted from Polar Kerr measurements (both rotation and ellipticity) 
as described in Ref.~\onlinecite{Ferreiro-Vila2009}. Let us emphasize 
that in the case of the Fe film, the permittivity tensor in the transversal 
configuration described in Fig.~\ref{scheme}(a) adopts the following 
form\cite{Zvezdin1997}
\begin{equation}
\label{perm-tensor-Fe}
\bar \epsilon = \left( \begin{array}{ccc}
\epsilon & 0 & \epsilon_{xz} \\
0 & \epsilon & 0 \\
- \epsilon_{xz} & 0 & \epsilon \end{array} \right) ,
\end{equation}
where $\epsilon$ is the permittivity function of the non-magnetized film and
$\epsilon_{xz} = a M$, where $M$ is the magnitude of the magnetization
at saturation. Let us recall that in the transverse configuration ${\bf M}
= M {\bf \hat y}$, \emph{i.e.}\ the magnetization is parallel to the Fe film
and perpendicular to the plane of incidence.

Since the holes of our structure are circular, the Fourier expansion of the 
permittivity, Eq.~(\ref{epG-Fourier}), can be calculated analytically.\cite{Plihal1991} 
For holes of radius $r$ and permittivity components $\epsilon^h_{ij}$ in a 
material with permittivity $\epsilon^m_{ij}$, we have
\begin{equation}
\tilde \epsilon_{ij}({\bf G}) = \left\{ \begin{array}{lr}
2(\epsilon^h_{ij} - \epsilon^m_{ij}) \beta J_1(Gr)/(Gr) & \mbox{if}\;\, {\bf G} \ne 0 \\
\epsilon^m_{ij} + \beta ( \epsilon^h_{ij} - \epsilon^m_{ij}) & \mbox{if}\;\, {\bf G} = 0 ,
\end{array} \right.
\end{equation}
where $\beta$ is the fraction of the area occupied by the holes, and $J_1$ is a Bessel
function of first kind. In the case of a triangular lattice with lattice constant $a_0$,
$\beta = (2/\sqrt{3}) \pi r^2/a^2_0$.

In the numerical calculations performed to obtain the results that we are about to
describe we have truncated the Maxwell equations by setting a high-momentum cutoff
and we have employed the fast Fourier factorization described in Appendix A. In
particular, the results shown in what follows were obtained by using $N_G = 367$
lattice vectors, which suffices to converge the different physical properties discussed
here (see Appendix A).

\begin{figure}[t]
\begin{center}
\includegraphics[width=\columnwidth,clip]{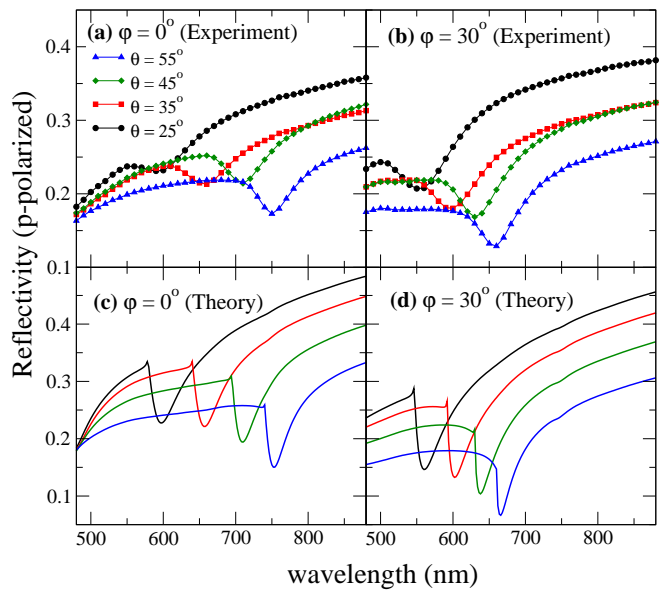}
\caption{(Color online) Experimental and theoretical results for the reflectivity of the
demagnetized structure along the $p$-channel, $R_{pp}$, as a function of the wavelength
of the incident light. As indicated in the panels, the results are shown for two different
high symmetry crystallographic directions $\varphi = 0^{\rm o}$ and $\varphi = 30^{\rm o}$
and for various angles of incidence $\theta$.}
\label{comp-R}
\end{center}
\end{figure}

Let us start our discussion of the results by describing the reflectivity in this
multilayer system when the Fe film is demagnetized. In the upper panels of
Fig.~\ref{comp-R} we reproduce the experimental results for $p$-polarized light
obtained for two different high symmetry crystallographic directions $\varphi =
0^{\rm o}$ and $\varphi = 30^{\rm o}$ and for various angles of incidence $\theta$
(see Fig.~\ref{scheme}(a) for a definition of these angles).\cite{note1} The most
prominent feature is the appearance of a dip which is red shifted as the angle
of incidence $\theta$ is increased. Notice that the red shift depends on the
crystallographic direction, and it is more pronounced for $\varphi = 0^{\rm o}$. 
Such a feature is absent in the case of $s$-polarized light (not shown here) and 
it can be attributed to the excitation of surface plasmon polaritons (SPPs), as 
we shall discuss below. In the lower panel of Fig.~\ref{comp-R}, we show the 
corresponding theoretical results calculated with the scattering approach assuming 
that $\epsilon_{xz}$ in Eq.~(\ref{perm-tensor-Fe}) is zero. As one can see, our 
calculations nicely reproduce the experimental trends. The theoretical dips appear 
to be more pronounced than in the experiment, which we attribute to the unavoidable 
inhomogeneities in the periodic array of holes in the Fe film.

\begin{figure}[t]
\begin{center}
\includegraphics[width=\columnwidth,clip]{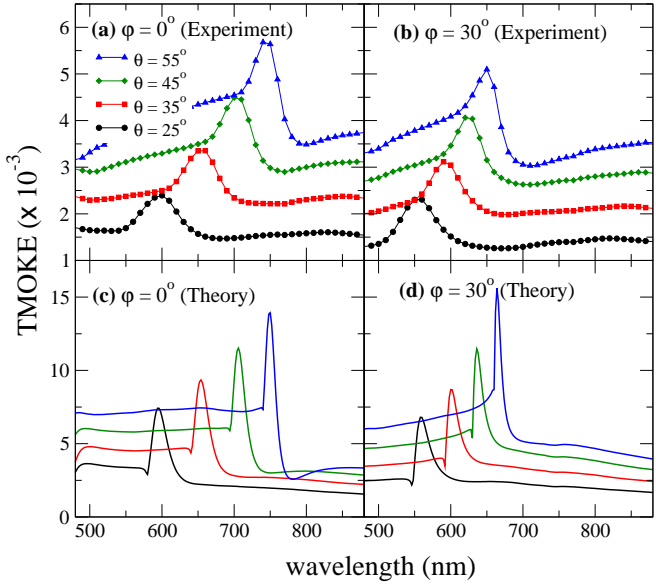}
\caption{(Color online) Experimental and theoretical results for the TMOKE as a
function of wavelength for $\varphi = 0^{\rm o}$ and $\varphi = 30^{\rm o}$ and
for various angles of incidence $\theta$.}
\label{comp-Tmoke}
\end{center}
\end{figure}

\begin{figure}[b]
\begin{center}
\includegraphics[width=0.8\columnwidth,clip]{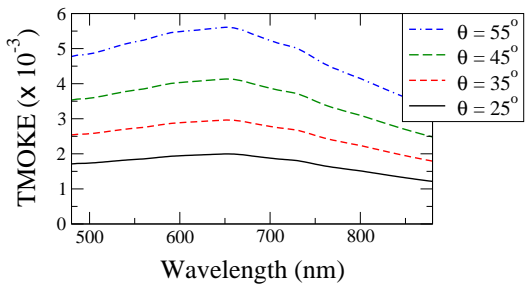}
\caption{(Color online) Theoretical results for the TMOKE for the non-perforated
multilayer structure (formed by uniform slabs) as a function of wavelength for
various angles of incidence $\theta$.}
\label{fe-layer}
\end{center}
\end{figure}

The corresponding results (both experimental and theoretical) for the TMOKE, as
defined in Eq.~(\ref{def-Tmoke}), are displayed in Fig.~\ref{comp-Tmoke}. Notice
that the theoretical results, in good agreement with the experiment, show that
the TMOKE can be resonantly enhanced at wavelengths that follow closely those
in which the dips in the reflectivity appear. Notice that at resonance the TMOKE
signal increases by roughly a factor of 2 with respect to value at off-resonant
wavelengths. It is worth stressing that, as illustrated in Fig.~\ref{fe-layer},
the signal for the continuous Fe film (no perforated) is featureless in the
spectral range considered here.

\begin{figure}[t]
\begin{center}
\includegraphics[width=\columnwidth,clip]{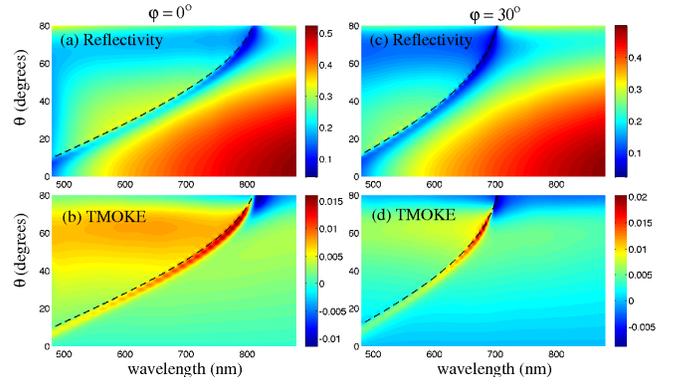}
\caption{(Color online) Theoretical results for the reflectivity and TMOKE as a
function of the wavelength and angle of incidence $\theta$ for $\varphi = 0^{\rm o}$
and $\varphi = 30^{\rm o}$. The dashed lines in the different panels correspond
to the resonant condition for the excitation of SPPs, as described by
Eq.~(\ref{ssp-disp}).}
\label{2D-plots}
\end{center}
\end{figure}

In order to understand the origin of the peaks in the TMOKE and the corresponding
dips in the reflectivity, we have investigated these quantities in a more
systematic way. In Fig.~\ref{2D-plots} we present the results for these two
quantities as a function of the wavelength and of the angle of incidence $\theta$.
In this figure we can observe again the appearance of the dips in the reflectivity,
which are accompanied by pronounced peaks in the TMOKE. The shape of the TMOKE and
the dispersion of the peaks with $\theta$ suggest that these features originate
from the excitation of the SPPs of this structure. To confirm this impression we
have to calculate the matching condition for the excitation of these surface modes.
For this purpose, we need first to determine the dispersion relation of the SPPs
and, as an approximation, we shall assume that it is given by the dispersion
relation for continuous films. Thus, ignoring the thin Au layer, which is practically
transparent, the complex wave vector of the SPP modes is given by\cite{Raether1988}
\begin{equation}
k_{\rm spp}(\lambda) = \frac{2\pi}{\lambda} \sqrt{\frac{\epsilon(\lambda)}
{1 + \epsilon(\lambda)}} ,
\end{equation}
where $\lambda$ is the light wavelength, $\epsilon$ is the permittivity of the
demagnetized Fe film, and we have used the fact that the incidence medium is air.
Now, the matching condition, which is the conservation of the parallel wave vector,
can be written as
\begin{equation}
\label{ssp-disp}
| \mbox{Re} \left\{ k_{\rm spp} (\lambda_{n_1,n_2}) \right\} | =
| {\bf k}_{||} + {\bf G}_{n_1,n_2} |,
\end{equation}
where ${\bf k}_{||} = (2\pi/\lambda) \sin \theta {\bf \hat x}$ is the in-plane 
wave vector of the incoming light (in the transverse configuration described in
Fig.~\ref{scheme}(a)) and ${\bf G}_{n_1 n_2}$ is a reciprocal lattice vector, 
which with our choice for the reciprocal lattice basis vectors (see 
Fig.~\ref{scheme}(b)) is given by ${\bf G}_{n_1,n_2} = (2\pi/a_0 \sqrt{3}) 
\{ [(2n_2 - n_1) \cos \varphi + n_1 \sqrt{3} \sin \varphi ] {\bf \hat x} + 
[ n_1 \sqrt{3} \cos \varphi + (n_1 -2n_2) \sin \varphi ] {\bf \hat y} \}$. 
The condition of Eq.~(\ref{ssp-disp}) tells 
us at which (discrete) wavelengths can the SPPs be excited for a given angle of
incidence. We have solved Eq.~(\ref{ssp-disp}) numerically and found that for
$\varphi = 0^{\rm o}$ the only mode that can be excited in the wavelength range
analyzed here is $\lambda_{0,-1}$, while for $\varphi = 30^{\rm o}$ we have
two possibilities: $\lambda_{0,-1}$ and $\lambda_{-1,-1}$, which indeed correspond
to the same wavelength. In Fig.~\ref{2D-plots} we have included as dashed lines
the relation between the resonant wavelength and the angle of incidence $\theta$
for these two cases. As one can see, these relations nicely describe the positions
of both the dips in the reflectivity and the peaks in the TMOKE. This strongly
suggests that the resonant enhancement of the TMOKE is due to the excitation of
SPPs in our layered structure.

\section{Conclusions} \label{sec-conclusions}

We have presented in this work a generalization of the scattering-matrix approach
to describe the propagation of electromagnetic waves in periodically patterned
multilayer structures containing materials with any kind of optical activity and
anisotropy. This generalized formalism enables us to tackle important physical problems
that have been traditionally out of the scope of this approach. Thus for instance,
the method can be applied to describe all the basic magneto-optical effects in any
possible configuration in magnetic structures. Moreover, the method can also be used
to study the wave propagation in periodic structures containing an arbitrary number
of birefringent/dichroic layers.

We have illustrated the use and capabilities of the method by analyzing a recent
experiment in which the transverse magneto-optical Kerr effect (TMOKE) was investigated
in a Fe film with a periodic array of subwavelength holes.\cite{Torrado2010} We have
shown, in excellent agreement with the experiment, that the TMOKE signal can be
resonantly enhanced when the samples are illuminated with an appropriate wavelength,
and we have attributed this phenomenon to the excitation of surface plasmon polaritons.
This resonant enhancement of the magneto-optical signal is closely related to the
phenomenon of extraordinary optical transmission (EOT),\cite{Garcia-Vidal2010} which
indeed takes place in these perforated Fe films. The systematic analysis of the interplay
between the EOT phenomenon and the different magneto-optical effects in perforated
magnetic films will be the subject of a forthcoming publication.

\section{Acknowledgements}

We would like to thank G. Armelles and J.F. Torrado  for providing us the 
experimental data on iron films, G.A. also provided the experimental values 
of the dielectric constants and E. Ferreiro-Vila performed the Polar Kerr 
experiments for the Fe film. BC and AG-M acknowledge funding from the EU
(NMP3-SL-2008-214107-Nanomagma), the Spanish MICINN (``MAPS''
MAT2011-29194-C02-01, ``MAGPLAS'' MAT2008-06765-C02-01/NAN and
``FUNCOAT'' CONSOLIDER INGENIO 2010 CSD2008-00023), the Comunidad
de Madrid (``MICROSERES-CM'' S2009/TIC-1476). JCC acknowledges financial 
support from the Spanish MICINN (Contract No.\ FIS2011-28851-C02-01) and 
from the German foundations Carl Zeiss Stiftung and Stifterverband f\"ur die 
Deutsche Wissenschaft via the award of an ``InnoLecture-Gastdozentur".

\appendix

\section{Fast Fourier factorization}

The scattering approach, as formulated in section \ref{SMA}, is known to have important
convergence problems when metals are involved. These problems are specially pronounced
when the structures contain noble metals and the infrared range is investigated. These
problems are well-known in the theory of gratings\cite{Neviere2003} and it has been
understood that they originate from the incorrect factorization of the product of two
periodic discontinuous functions. Such a product appears, in particular, in the
constitutive relation ${\bf D} = \bar \epsilon {\bf E}$, where ${\bf D}$ is the
displacement vector. When calculating the Fourier components of ${\bf D}$ in section
\ref{subsec-maxwell}, we have used the so-called Laurent's rule. This rule states that
the Fourier components $h_n$ of the product $h(x)$ of two arbitrary functions $f(x)$
and $g(x)$ are given by
\begin{equation}
h_n = \sum^{\infty}_{m=-\infty} f_{n-m} g_m .
\end{equation}
Although this result is correct, as long as the sum extends to infinity, it is not
always correct when one truncates the series, as we do numerically. This was recognized
by Li,\cite{Li1996} who established the following rules for factorization:
\begin{enumerate}
\item Let $h(x) = f(x) g(x)$ and either $f(x)$ or $g(x)$ be continuous at some $x=x_0$.
The other quantity may be discontinuous there. Then, Laurent's rule applies, {\it i.e.}\
\begin{equation}
\left[ h \right] = \left[ \left[ f \right] \right] \left[ g \right] .
\end{equation}
Here, $[g]$ denotes a column vector constructed with, let us say, $N_G$ Fourier components
$g_n$ and by $[[f]]$ we denote the $N_G \times N_G$ Toeplitz matrix whose $(n,m)$ entry is
$f_{n-m}$.

\item Let $h(x) = f(x) g(x)$ and both $f(x)$ and $g(x)$ be discontinuous at some $x=x_0$,
but the product $f(x)g(x)$ be continuous there. Then, the so-called inverse rule holds,
which is given by
\begin{equation}
\left[ h \right] = \left[ \left[ \frac{1}{f} \right] \right]^{-1} \left[ g \right] .
\end{equation}

\item Let $h(x) = f(x) g(x)$ and both $f(x)$ and $g(x)$ be discontinuous at some $x=x_0$
and the product $f(x)g(x)$ be discontinuous there as well. Then, the product of the two
functions in Fourier space cannot be formed by either the Laurent's rule or the inverse
rule.
\end{enumerate}

Obviously, in our analysis of the Maxwell's equations in section \ref{subsec-maxwell},
see Eqs.~(\ref{Amp1}-\ref{Amp3}), we are violating these factorization rules. We are
simply using the Laurent's rule, although in the interface between different materials
we may have concurrent discontinuities in both the permittivity tensor and the electric
field, and in some cases the product (the displacement vector) is discontinuous as well.
Thus, our goal now is to reformulate the Maxwell equations in momentum space in order to
respect the factorization rules stated above. For this purpose, we make use of the
so-called \emph{fast Fourier factorization} put forward by Popov and Nevi\`ere in
Ref.~\onlinecite{Popov2001}.

For the sake of concreteness, let us consider a two-dimensional periodic system consisting
of an array of circular holes or circular pillars, as in the structure of section
\ref{sec-antidot}. Now, let us define a vector with the continuous components of the
${\bf E}$ and ${\bf D}$ fields, {\it i.e.}\ ${\bf G} = [E_t, D_n, E_z]^T$. Here, $E_t$
is the tangential component of the electric field in the $xy$ plane, $D_n$ is the normal
component of the displacement vector in the $xy$ plane, and $E_z$ is the $z$ component of
the electric field. These three components are continuous in the $xy$ plane when we cross
the boundary of a hole (or pillar) and the permittivity tensor undergoes a discontinuity.
Now, let us establish the relation between these field components and the three Cartesian
components of the electric field ${\bf G} = \hat F {\bf E}$, where ${\bf E} = [E_x, E_y,
E_z]^T$. There are many possible choices for $\hat F$. We choose to express its matrix
elements in terms of the polar angle $\phi(x,y)$ defined as $r e^{i\phi(x,y)} = x+iy$.
It is straightforward to show that
\begin{equation}
\hat F = \left( \begin{array}{ccc} -s & c & 0 \\
\epsilon_{xx} c + \epsilon_{yx} s & \epsilon_{xy} c + \epsilon_{yy} s &
\epsilon_{xz} c + \epsilon_{yz} s \\ 0 & 0 & 1 \end{array} \right),
\end{equation}
where $c$ and $s$ are abbreviations for $\cos \phi$ and $\sin \phi$, respectively.

We now define the inverse of this matrix $\hat C = \hat F^{-1}$. Thus, ${\bf E}
= \hat C {\bf G}$. Let us recall the constitutive relation ${\bf D} = \bar
\epsilon {\bf E}$, where $\bar \epsilon$ is given by Eq.~(\ref{perm-tensor}).
This relation can be now written as
\begin{equation}
{\bf D} = \bar \epsilon \hat C \cdot {\bf G} = \bar \epsilon \hat C \cdot
\hat F {\bf E} ,
\end{equation}
whose elements are now expressed as the product of a discontinuous function and a
continuous one. Thus, using Laurent's rule for the first term of the product
and the inverse rule for the second one, the Fourier components of the displacement
vector can be calculated as
\begin{equation}
[{\bf D}] = [[ \bar \epsilon \hat C]] \, [[\hat C ]]^{-1} [{\bf E}] .
\end{equation}
This indicates that the Toeplitz matrix of the index tensor $\hat \eta$ in the
formalism of section \ref{SMA} has to be calculated as follows
\begin{equation}
\label{eta-Toeplitz}
[[ \hat{\hat \eta} ]] = [[ \hat C ]] [[ \bar \epsilon \hat C ]]^{-1} .
\end{equation}
This is indeed the only change that we need to introduce in the formalism to improve
significantly the convergence in the problematic cases. Notice that in practice this
requires the calculation of the Toeplitz matrix of several trigonometric functions,
which in general has to be done numerically.

For the sake of completeness, we now provide simplified expressions for $[[\hat{\hat \eta}]]$
in some cases of special interest for us. First, in the case of an isotropic material,
for which $\bar \epsilon = \epsilon \hat 1$, it is straightforward to show that
Eq.~(\ref{eta-Toeplitz}) reduces to\cite{David2006,Antos2010}
\begin{eqnarray}
\left[ \left[ \hat{\hat \eta} \right] \right] & = & \\ & & \hspace*{-1.2cm}
\left( \begin{array}{ccc}
\left[ \left[ \epsilon \right] \right]^{-1} + \left[ \left[ X \right] \right]
\left[ \left[ c^2 \right] \right] & \left[ \left[ X \right] \right]
\left[ \left[ c s \right] \right] & 0 \\
\left[ \left[ X \right] \right] \left[ \left[ c s \right] \right] &
\left[ \left[ 1/ \epsilon \right] \right] - \left[ \left[ X \right] \right]
\left[ \left[ c^2 \right] \right] & 0 \\ 0 & 0 & \left[ \left[ \epsilon
\right] \right]^{-1} \end{array} \right) , \nonumber
\end{eqnarray}
where $[[X]] = [[1/\epsilon]] - [[\epsilon]]^{-1}$. This requires, in particular,
the calculation of the Fourier components of the trigonometric functions $\cos^2 \phi$
and $\cos \phi \sin \phi$, which can be easily done numerically. Notice that these
components are independent of the wavelength of the light and therefore, they can be
calculated once and forever for a given structure. On the other hand, for the description
of the TMOKE we have to consider a permittivity tensor given by Eq.~(\ref{perm-tensor-Fe}).
In this case, the Fourier components of the index tensor are given by
\begin{widetext}
\begin{equation}
\left[ \left[ \hat{\hat \eta} \right] \right]^{-1} = \left( \begin{array}{ccc}
\left[ \left[ \epsilon \right] \right] + \left[ \left[ Y \right] \right]
\left[ \left[ c^2 \right] \right] & \left[ \left[ Y \right] \right]
\left[ \left[ c s \right] \right] & \left[ \left[ \epsilon_{xz}
\right] \right] + \left[ \left[ Z^{\prime} \right] \right]
\left[ \left[ c^2 \right] \right] \\
\left[ \left[ Y \right] \right] \left[ \left[ c s \right] \right] &
\left[ \left[ 1/\epsilon \right] \right]^{-1} - \left[ \left[ Y \right] \right]
\left[ \left[ c^2 \right] \right] &  \left[ \left[ Z^{\prime} \right] \right]
\left[ \left[ c s \right] \right] \\
- \left[ \left[ \epsilon_{xz} \right] \right] - \left[ \left[ Z \right] \right]
\left[ \left[ c^2 \right] \right] & - \left[ \left[ Z \right] \right]
\left[ \left[ c s \right] \right] & \left[ \left[ \epsilon \right] \right] +
\left[ \left[ W \right] \right] \left[ \left[ c^2 \right] \right] \end{array} \right) ,
\end{equation}
\end{widetext}
where
\begin{eqnarray}
[[Y]] & = & [[1/\epsilon]]^{-1} - [[\epsilon]], \nonumber \\
\left[ \left[ Z \right] \right] & = & [[\epsilon_{xz}/ \epsilon]] \,
[[1/\epsilon]]^{-1} - [[\epsilon_{xz}]], \nonumber \\
\left[ \left[ Z^{\prime} \right] \right] & = & [[1/\epsilon]]^{-1}
[[\epsilon_{xz} / \epsilon]] - [[\epsilon_{xz}]], \nonumber \\
\left[ \left[ W \right] \right] & = & [[ \epsilon^2_{xz} / \epsilon]] -
[[\epsilon_{xz}/ \epsilon]] \, [[1/\epsilon]]^{-1} [[\epsilon_{xz}/ \epsilon]].
\end{eqnarray}
Again, one just needs the evaluation of the Fourier components of both $\cos^2 \phi$
and $\cos \phi \sin \phi$.

It is worth stressing that in the choice of the normal vectors entering in the Fourier
factorization there is a freedom that one can use to further improve the convergence
of the calculations. For a discussion of this issue, see Ref.~\onlinecite{Antos2010}.

\begin{figure}[t]
\begin{center}
\includegraphics[width=\columnwidth,clip]{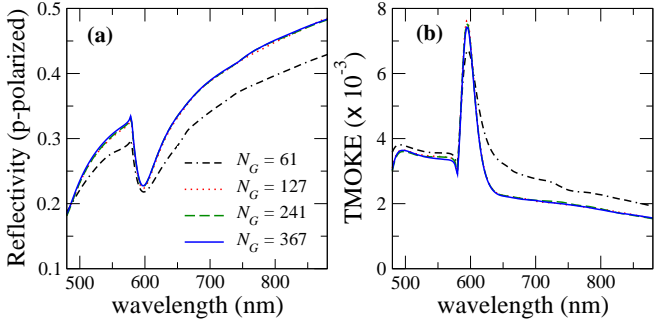}
\caption{(Color online) Reflectivity in the absence of magnetic field (a) and TMOKE
(b) for the multilayer structure studied in section \ref{sec-antidot} for $\theta =
25^{\rm o}$ and $\varphi = 0^{\rm o}$ as a function of the wavelength of the incident
light. The different curves correspond to results obtained with different number of
reciprocal lattice vectors $N_G$.}
\label{convergence}
\end{center}
\end{figure}

To conclude, we now want to illustrate the convergence of the results using the fast
Fourier factorization described in this appendix. In Fig.~\ref{convergence} we show an
example of the results obtained for the reflectivity and the TMOKE for the structure
studied in section \ref{sec-antidot}. In this figure, the different curves correspond
to different values of $N_G$, which is the number of reciprocal lattice vectors taken
into account in the calculations upon setting a high-momentum cutoff. As one can see,
it is possible to converge the calculations to a high precision in the whole range
of wavelengths. Moreover, the convergence is rapid and uniform. It is important to
emphasize that in order to get results of similar quality for this example without
the use of the fast Fourier factorization, values of $N_G$ even larger than 1000 are
required (not shown here).

\section{Solving the nonlinear eigenvalue problem}

We detail here a simple strategy to solve numerically the nonlinear eigenvalue problem
of Eq.~(\ref{rational-eigen}), which we have found to work without any problem in all
the cases that we have considered. The first step is to multiple both sides of
Eq.~(\ref{rational-eigen}) by $q$ to convert it into the following cubic eigenvalue
problem
\begin{equation}
\left( {\cal B}_3 q^3 + {\cal B}_2 q^2 + {\cal B}_1 q + {\cal B}_0
\right) \phi = 0 ,
\label{cubic-eigen}
\end{equation}
where we have defined ${\cal B}_n = {\cal A}_{n-1}$. Now, the simplest strategy
to solve this cubic problem is to use a standard linearization procedure.\cite{Gohberg1982}
The idea goes as follows. We first define the following vectors
\begin{equation}
\lambda_n = q^{n-1} \phi; \;\;\; n=1,2,3.
\end{equation}
From this definition, and using Eq.~(\ref{cubic-eigen}), it is easy to show that
the vectors $\lambda_n$ satisfy the following equation
\begin{equation}
\left( \begin{array}{ccc} 0 & 1 & 0 \\ 0 & 0 & 1 \\
{\cal B}_0 & {\cal B}_1 & {\cal B}_2 \end{array} \right)
\left( \begin{array}{c} \lambda_1 \\ \lambda_2 \\ \lambda_3 \end{array} \right) =
q \left( \begin{array}{ccc} 1 & 0 & 0 \\ 0 & 1 & 0 \\
0 & 0 & -{\cal B}_3 \end{array} \right) \left( \begin{array}{c} \lambda_1 \\ \lambda_2
\\ \lambda_3 \end{array} \right) .
\end{equation}
We have thus converted the problem into a generalized linear eigenvalue problem that can
be solved with standard linear algebra techniques. The obvious disadvantage of this simple
procedure is that one increases the dimension of the problem by a factor of 3. In this
sense, it may be advantageous in some cases to implement other methods like, for instance,
the iterative Newton method. In any case, and as illustrated in the previous appendix, we have
not found problems to converge the calculations detailed in this work with the linearization
procedure.

\section{Spatially uniform slabs}

A multilayer structure may contain some uniform (non-structured) layers. In particular,
this is always the case for the medium of incidence and for the substrate layer. In
this sense, it is interesting to discuss how the formalism discussed in section
\ref{SMA} is simplified in the case of uniform slabs. In this case, the permittivity
tensor is diagonal in momentum space: $(\hat \epsilon_{ij})_{{\bf G},{\bf G}^{\prime}}
= \tilde \epsilon_{ij}(0) \delta_{{\bf G},{\bf G}^{\prime}}$. This implies that all
the matrices in momentum representation are also diagonal. The eigenvalue problem of
Eq.~(\ref{rational-eigen}) leads to the following quartic secular equation for
$q({\bf G})$. Focusing on ${\bf G} = 0$, this equation reads
\begin{equation}
\label{sec-eq}
\sum^4_{n=0} D_n q^n = 0 ,
\end{equation}
where the coefficients are given by
\begin{eqnarray}
D_4 & = & \eta_{xx} \eta_{yy} - \eta_{xy} \eta_{yx} ,
\nonumber \\
D_3 & = & k_x \left[ n_{xy} \eta_{yz} + \eta_{yx} \eta_{zy} - n_{yy}
(\eta_{xz} + \eta_{zx}) \right] \nonumber \\ &&  + k_y \left[ n_{yx}
\eta_{xz} + \eta_{xy} \eta_{zx} - n_{xx} (\eta_{yz} + \eta_{zy}) \right] ,
\nonumber \\
D_2 & = & k^2_x \left[ \eta_{yy} (\eta_{xx} + \eta_{zz}) - \eta_{xy} \eta_{yx}
- \eta_{yz} \eta_{zy} \right] \nonumber \\ && + k^2_y \left[ \eta_{xx} (\eta_{yy} + \eta_{zz}) -
\eta_{xy} \eta_{yx} - \eta_{xz} \eta_{zx} \right] \nonumber \\ & & + k_x k_y \left[
\eta_{xz} (\eta_{yz} + \eta_{zy}) + \eta_{yz} (\eta_{zx} - \eta_{xz}) \right.
\nonumber \\ && - \left. \eta_{zz} (\eta_{xy} + \eta_{yx}) \right] - \omega^2
(\eta_{xx} + \eta_{yy}) ,
\nonumber \\
D_1 & = & k^3_x \left[ \eta_{xy} \eta_{yz} + \eta_{yx} \eta_{zy} - \eta_{yy}
(\eta_{xz} + \eta_{zx}) \right] \nonumber \\ && + k^3_y \left[ \eta_{yx} \eta_{xz} +
\eta_{xy} \eta_{zx} - \eta_{xx} (\eta_{yz} + \eta_{zy}) \right] \nonumber \\ & &
+ k^2_x k_y \left[ \eta_{xy} \eta_{zx} + \eta_{xz} \eta_{yx} - \eta_{xx}
(\eta_{yz} + \eta_{zy}) \right] \nonumber \\ && + k^2_y k_x
\left[ \eta_{yx} \eta_{zy} + \eta_{yz} \eta_{xy} - \eta_{yy} (\eta_{xz} + \eta_{zx})
\right] \nonumber \\ & & + \omega^2 \left[ k^2_x(\eta_{xz} + \eta_{zx}) + k^2_y
(\eta_{yz} + \eta_{zy}) \right] ,
\nonumber \\
D_0 & = & k^4_x (\eta_{yy} \eta_{zz} - \eta_{yz} \eta_{zy}) + k^4_y (\eta_{xx}
\eta_{zz} - \eta_{xz} \eta_{zx}) \nonumber \\ & &
+ k^3_x k_y \left[ \eta_{xz} \eta_{zy} + \eta_{yz} \eta_{zx} - \eta_{zz}
(\eta_{xy} + \eta_{yx}) \right] \nonumber \\ & &
+ k^3_y k_x \left[ \eta_{yz} \eta_{zx} + \eta_{xz} \eta_{zy} - \eta_{zz}
(\eta_{yx} + \eta_{xy}) \right] \nonumber \\ & &
+ k^2_x k^2_y \left[ \eta_{zz} (\eta_{xx} + \eta_{yy}) + \eta_{xy} \eta_{yx} -
\eta_{xz} \eta_{zx} - \eta_{yz} \eta_{zy} \right] \nonumber \\ & &
+ \omega^2 \left[ \omega^2 - k^2_x ( \eta_{yy} + \eta_{zz}) - k^2_y (\eta_{xx} + \eta_{zz})
\right. \nonumber \\ & &
\hspace*{8mm} + \left. k_x k_y (\eta_{xy} + \eta_{yx}) \right] .
\end{eqnarray}
For ${\bf G} \ne 0$, one just needs to replace $k_{x,y}$ by $k_{x,y} + G_{x,y}$.
Eq.~(\ref{sec-eq}) has been previously derived (in terms of the components of the
permittivity tensor) in the context of the analysis of uniform multilayer structures 
containing magneto-optical and anisotropic materials.\cite{Mansuripur1990} This 
equation simplifies in several limiting cases. Thus for instance, if we consider 
the typical configuration for measuring the TMOKE, then the permittivity tensor 
is given by Eq.~(\ref{perm-tensor-Fe}). In this case, $D_3 = D_1 = 0$ and setting 
$k_y=0$ the solutions of Eq.~(\ref{sec-eq}) are $q^2_1 = \omega^2/\eta_{yy} - k^2_x$
and $q^2_2 = \omega^2/\eta_{xx} - k^2_x$. Moreover, in this case the layer matrix 
defined in Eq.~(\ref{M-def}) adopts the following simple form (for ${\bf G} = 0$)
\begin{equation}
M = \left( \begin{array}{cccc}
\omega^2/q_1 & 0 & -\omega^2/q_1 & 0 \\
0 & \eta_{xx} q_2 - \eta_{xz} k_x & 0 & -\eta_{xx} q_2 - \eta_{xz} k_x \\
1 & 0 & 1 & 0 \\ 0 & 1 & 0 & 1 \end{array} \right) .
\end{equation}
On the other hand, for an isotropic layer $\bar \epsilon = \epsilon \hat 1$, 
and in this case $D_3 = D_1 = 0$ and $q^2 = \epsilon \omega^2 -(k^2_x + k^2_y)$. 
The corresponding layer matrix reads now (for ${\bf G} = 0$)
\begin{equation}
M = \left( \begin{array}{cc}
M_{11} & -M_{11} \\
\hat 1 & \hat 1 \end{array} \right) ,
\end{equation}
where $\hat 1$ is the $2 \times 2$ unit matrix and
\begin{equation}
M_{11} = \frac{1}{q} \left( \begin{array}{cc}
\omega^2-k^2_y \eta & \eta k_x k_y \\
\eta k_x k_y & \omega^2-k^2_x \eta \end{array} \right) .
\end{equation}
%



\begin{thebibliography}{10}
\bibitem{note1}
Let us stress that $\varphi$ is the angle by which the sample is rotated
about the $z$-axis, see Fig.~\ref{scheme}(a), while the plane of incidence and 
the magnetic field are kept fixed and perpendicular to each other.
\end{thebibliography}
\end{document}